\documentclass[12pt]{article}
\usepackage{amsmath}
\usepackage{amssymb}
\usepackage{amsfonts}
\renewcommand{\theequation}{\arabic{section}.\arabic{equation}}

\begin{document}

\thispagestyle{empty}\mbox{}\setcounter{page}{0}\noindent

\begin{center}
{\LARGE \textbf{The Null Decomposition of Conformal Algebras}\bigskip\ 
\vspace{10pt}\\[0pt]
}

\bigskip {\LARGE \ }

{\large Dana Mihai, George A.J. Sparling\\[0pt]
Department of Mathematics\\[0pt]
University of Pittsburgh}

{\large Pittsburgh, Pennsylvania, USA\bigskip \bigskip\ \vspace{10pt}}

{\large\textbf{Abstract}}\\\vspace{5pt}
\begin{quote}\textbf{We analyze the decomposition of the enveloping algebra of
the conformal algebra in arbitrary dimension with respect to the
mass-squared operator. It emerges that the subalgebra that commutes with the mass-squared is generated by its Poincar\'{e} subalgebra together with a vector operator. The special cases of the
conformal algebras of two and three dimensions are described in detail,
including the construction of their Casimir operators.}\end{quote}
\end{center}

\pagebreak 

\section{Introduction}

The study of symmetry groups has shown that invariants are of great interest
in many areas in mathematics and physics, being used not only to label
irreducible representations of Lie algebras, and decomposing generic
representations into irreducible ones, but also in studying symmetry
breaking [9]. (Symmetry breaking refers to systems which possess a certain
symmetry and are perturbed in such a way that only part of the original
symmetry still holds in the system. More often than not, the perturbed
system presents more interest than the original one, since the full symmetry
in general describes situations which lack reality.) In this process, one
can still relate the invariants of the whole symmetry group with those of
its subgroups [2].\medskip

The Lie algebra $\mathfrak{su}(2)$ played an important role in studying the
mass spectrum of hadrons in particle physics. The central idea is that
particles with the same spin, parity and close mass values should be related
by some kind of symmetry [3]. In 1932, Heisenberg regarded the neutron and
proton as forming basis states for the fundamental representations of $%
\mathfrak{su}(2)$ [5]. One of the quantum numbers used to label the
particles is the isotopic spin (or isospin), which is conserved in reactions
involving electromagnetic and strong interactions. This is not the only
quantum number conserved in these reactions. In order to accomodate this
extra quantity, in 1964 Gell-Mann and Ne'eman [6] grouped the hadrons into
multiplets of $\mathfrak{su}(3)$. The basic observed particles fitted though
into an eight-dimensional adjoint representation instead of the fundamental
three-dimensional $\mathfrak{su}(3)$ representation.\medskip

A natural question to ask (from the point of view of supersymmetry) is
whether there is some symmetry connnecting the different octets in this
eight-dimensional adjoint representation. In [1], Sparling suggested that
such a symmetry arises naturally in the context of broken conformal
invariance; there he addressed the question of what happens to a system
transforming under the conformal group when the conformal invariance is
broken by fixing the mass.\medskip

In the absence of gravitation, the physical space-time is the Minkowski
space $\mathbb{M}$, which is a four-dimensional real vector space with a
metric of signature $(3,1)$ (or $(1,3)$). As a manifold, this space is
isomorphic to its group of translations. The transformations under the
Lorentz group $\mathbb{SO}(3,1)$ of rotations of $\mathbb{M}$ contain also
important information on the physics of the system.

In order to include them in the formalism, $\mathbb{M}$ has to be regarded
as the homogeneous space of the Poincar\'{e} group, modulo the Lorentz group
[3]. Many techniques are created for semisimple Lie groups, but
unfortunately the Poincar\'{e} group is not semisimple. One solution that
would allow using these techniques is to regard a suitable compactification
of $\mathbb{M}$ as a coset of the group of conformal transformations of $%
\mathbb{M}$. \medskip

Physicists use the conformal group especially because it is simple. The
downside is that its action on a quantum mechanical system changes its mass
continuously. The Poincar\'{e} group, used extensively in particle physics
and quantum field theory (QFT), is the part of the conformal group that
commutes with the operator representing the square of the mass.\medskip

Quantum mechanics assumes that operators on a quantum system may be commuted
with each other, and also multiplied together in an associative way. For a
Lie algebra acting on a quantum system, this means that the action extends
naturally to an action of the enveloping algebra of its Lie algebra. (The 
\textit{enveloping algebra} of a given algebra $A$ is the smallest
associative algebra that contains $A$ as a subspace and for which the
commutator on this subspace reproduces the Lie bracket in $A$).\medskip

Therefore, from a QFT perspective, it is natural to study the subalgebra of
the enveloping algebra of the conformal algebra that commutes with the mass.
We find that this subalgebra, denoted $\mathcal{R}$, is strictly larger than
the enveloping algebra of the Poincar\'{e} Lie algebra; the extra structure
is encoded in a vector operator $R_{a}$ which is shown to commute with the
mass of the system and with the translations $p_{a}$. When the conformal
invariance is broken, provided the operator $R_{a}$ survives, no useful
information is lost, only the dilation operator $D$ needs to be removed. The
full conformal algebra is simply obtained by adding the generator \thinspace 
$D$ with appropriate commutation relations to the $\mathcal{R}$-algebra.
\medskip

We should remark here that the newly obtained $\mathcal{R}$-algebra,
although finitely generated, is not a finite-dimensional Lie-algebra, as the
commutator of the vector operator $R_{a}$ with itself is non-linear
(actually cubic) in the generators.\medskip

For a finite-dimensional simple algebra $A$, the universal enveloping
algebra $U$ contains a set of operators which can be used to label the
representations of $A$. These operators, the \textit{Casimir invariants},
generate the center of the associative algebra $U(A)$ and commute with any
of its elements [3], [4].\pagebreak

In this work, we are interested in constructing the $\mathcal{R}$-algebras
for spaces of various dimensions and their Casimir invariants. The
four-dimensional case has been thoroughly studied in [1]; its origins lie in
the study of the twistor model of hadrons which seeks to describe a hadron
by means of three twistors [7], [8].   These types of representations (for $n=4$) have also been used in the AdS/CFT correspondence [9], [10], [11].

The structure of the present work is as follows: in section two we generalize
the four-dimensional theory to arbitrarily many dimensions and signatures. In section three we
independently develop the three-dimensional theory and verify the consistency of
these results with the generalized case. Section four treats the two-dimensional
case, but calculations are only carried in the $(2,0)$ signature.\medskip

There are many intermediate formulas we used to derive the final expressions
included in this paper. Due to the fact that calculations are extremely
lengthy, we listed most of the formulas used in an appendix.\medskip

Although this theory a priori only applies to the (non-compact)
pseudo-orthogonal groups, many other (non-compact) Lie groups such as the
pseudo-unitary groups are subgroups of the pseudo-orthogonal group in a
natural way. As a consequence, this theory can be applied to these groups
also, with appropriate modifications.\medskip

Finally note that the de Sitter algebra of $n$-dimensional de-Sitter space, $\mathbb{SO}(1, n)$ and the anti-de-Sitter algebra of $n$-dimensional anti-de-Sitter space, $\mathbb{SO}(2, n-1)$ both fall within the scope of our theory [9].

\pagebreak

\section{The Conformal Algebra in $n$-dimensions\protect\bigskip \protect%
\medskip}
\subsection{Basic Commutation Relations\protect\medskip \protect\medskip}
\setcounter{equation}{0}
We consider the conformal algebra of an $n$-dimensional affine space
equipped with a metric of signature $(p,q)$, where $p$ and $q$ are
non-negative integers such that $p+q=n$. The algebra in question is the Lie
algebra of $\mathbb{SO}(p+1,q+1)$.\bigskip

Explicitly, we may write the following commutation relation:%
\begin{equation}\label{2.1}
\lbrack M_{AB},M_{CD}]=g_{AC}M_{BD}-g_{BC}M_{AD}-g_{AD}M_{BC}+g_{BD}M_{AC}.
\end{equation}
Here the upper case Latin indices run from $0$ to $n + 1$.  Also the Lie algebra generators $M_{AB}$ are skew: $M_{AB} = - M_{BA}$.  The metric $g_{AB} = g_{BA}$ is a real symmetric metric of signature $(p + 1, q + 1)$.   
We use a null decomposition of the metric $g_{AB}$, describing it by the $(n+2)\times (n+2)$ symmetric
matrix:%
\begin{equation}\label{2.2}
g_{AB}=\left( 
\begin{array}{ccc}
0 & 0 & 1 \\ 
0 & g_{ab} & 0 \\ 
1 & 0 & 0%
\end{array}%
\right) . 
\end{equation}%
So here the lower case Latin index runs from $1$ to $n$.  Also we have the components $g_{00} = g_{n+1 n + 1} = 0$, $g_{0a} = g_{a0} = g_{n+1 a} = g_{a n +1} = 0$.  The $n \times n$-symmetric matrix $g_{ab} = g_{ba} $ has signature $(p, q)$.     Now put 
\begin{equation}
M_{0a}=p_{a},\ M_{n+1a}=q_{a},\ M_{0n+1}=D, \textrm{and }M_{ab}=-M_{ba}, 
\label{2.3}
\end{equation}%
So the tensor $M_{AB}=-M_{BA}$ can be described by an $(n+2)\times (n+2)$ skew
matrix:

\begin{equation}\label{2.4}
M_{AB}=\left( 
\begin{array}{ccc}
0 & p_{a} & D \\ 
-p_{b} & M_{ab} & -q_{b} \\ 
-D & q_{a} & 0
\end{array}
\right) ,  
\end{equation}
$M_{ab}$ is itself an $n\times n$ skew matrix and represents the analogue of
the angular momentum in $n$ dimensions, $p_{a}$ represents the
translations,\ $q_{a}$ the special conformal transformations, and $D$ is the
dilation operator.\smallskip \medskip

The generators of the full algebra are thus: $p_{a}$,\ $q_{a}$,\ $M_{ab}$
and $D$.\bigskip \medskip
\noindent \eject
From equations (2.1)-(2.4), we obtain the following basic commutation
relations:%
\[
\lbrack p_{a},p_{b}]=0,\ \ \ [q_{a},q_{b}]=0,\ \ \
[p_{a},q_{b}]=M_{ab}+g_{ab}D, 
\]%
\begin{equation}
\lbrack M_{ab},p_{c}]=g_{ac}p_{b}-g_{bc}p_{a},\ \ \ \
[M_{ab},q_{c}]=g_{ac}q_{b}-g_{bc}q_{a},\medskip  \label{2.5}
\end{equation}%
\[
\lbrack
M_{ab},M_{cd}]=g_{ac}M_{bd}-g_{bc}M_{ad}-g_{ad}M_{bc}+g_{bd}M_{ac},\medskip 
\]%
\[
\lbrack D,p_{a}]=-p_{a},\ \ \ [D,q_{a}]=q_{a},\ \ \
[D,M_{ab}]=0. 
\]%
\medskip

\noindent We will consider the following subalgebras, satisfying the
commutation relations (2.5):\smallskip

\begin{itemize}
\item the full algebra $\mathcal{C}$, spanned by $p_{a},\ M_{ab},\ D,\ $and $%
q_{a}$, of real dimension $(n+1)(n+2)/2$;

\item the subalgebra \ $\mathcal{D}$ of $\mathcal{C}$, spanned by $p_{a},\
M_{ab},\ $and $D$, of real dimension $(n^{2}+n+2)/2$; this is called the 
\textit{Weyl algebra};

\item the subalgebra $\mathcal{P}$ of $\mathcal{C}$, spanned by $p_{a},\ $%
and $M_{ab}$, of real dimension $n(n+1)/2$; this is called the \textit{%
Poincar\'{e} algebra}, or the \textit{Euclidean algebra}, depending on the
signature of the metric.
\end{itemize}

Following the notation introduced in [1], we denote by $\mathcal{C}_{e},\ 
\mathcal{D}_{e},\ \mathcal{P}_{e}$ their corresponding enveloping
algebras.\medskip

Throughout this paper we will use $p^{2}$ or $p\cdot p$ to denote $%
p_{a}p^{a} $, and also $p^{-2}= (p^2)^{-1} = (p\cdot p)^{-1}=(p_{a}p^{a})^{-1}$. Adjoin to
each of the above algebras the element $p^{-2}$ and its powers $(p^{-2})^r$,
where $r$ is a positive integer, satisfying:
\begin{equation}
p^{-2}p^{2}=p^{2}p^{-2}=1, \label{2.6}
\end{equation}
and 
\begin{equation}
\lbrack p^{-2},X]=-p^{-2}[p^{2},X]p^{-2}, \label{2.7}
\end{equation}
for any $X$ in $\mathcal{C}_{e}$. Note that $p^{-2}$ is the inverse of $%
p^{2} $, where $p^{2}$ is the mass-squared operator of the system. The
resulting algebras will be denoted by $\mathcal{C}_{e}\left( p^{2}\right) ,$ 
$\mathcal{D}_{e}\left( p^{2}\right) ,$ $\mathcal{P}_{e}\left( p^{2}\right) $%
, so implicitly the representations that this work applies to are those of
non-zero mass only. It is well known that $p^{2}$ is a Casimir invariant for 
$\mathcal{P}_{e}$.\bigskip

\subsection{The $\mathcal{R}$-algebra}

The first step in our analysis is the introduction of an operator
constructed in $\mathcal{P}_{e}\left( p^{2}\right) $ that behaves like a
position operator:%
\begin{equation}
2p^{2}x_{a}=  M_{ab}p^{b}+p^{b}M_{ab} = 2x_ap^2,  \label{2.8}
\end{equation}%
satisfying the relation:%
\begin{equation} \label{2.8a}
x\cdot p=\frac{n-1}{2}. \end{equation}
Note that $x_{a}$ is not a full position operator since it obeys the
constraint relation (2.9). Also note that we can write $x_{a}$ as: 
\begin{equation}
p^{2}x_{a}=M_{ab}p^{b}+kp_{a},  \label{2.9}
\end{equation}%
where we abbreviate the quantity $\frac{n-1}{2}$ by $k$: so we have $k=x\cdot p=\frac{n-1}{2}$.  The operators $x_{a}$ and $p^{2}$ commute since $x_{a}$
lies in $\mathcal{P}_{e}\left( p^{2}\right) $.\medskip

Define also the analogue of the orbital angular momentum in $n$ dimensions:%
\begin{equation}
L_{ab}=x_{a}p_{b}-x_{b}p_{a}  \label{2.10}
\end{equation}%
and the intrinsic spin:%
\begin{equation}
S_{ab}=M_{ab}-L_{ab} = - S_{ba}.  \label{2.11}
\end{equation}
It is now straightforward to verify that $S_{ab}$ is orthogonal to $p_a$:
\begin{equation} S_{ab}p^b = p^bS_{ab} = 0.\label{2.11a}
\end{equation}
Finally define a projected metric tensor: $h_{ab}=g_{ab}-p^{-2}p_{a}p_{b}$.
Note that $p^{a}h_{ab}=h_{ab}p^{a}=0$ and $h_{a}^{\hspace{4pt}a}=n-1$. The
commutation relations satisfied by these new operators can be found in
appendix A.\medskip

The algebra $\mathcal{P}_{e}(p^{2})$ can be written now in terms of the
operators $x_{a},\ p_{a},\ S_{ab}$ (whose number of degrees of freedom are $n-1,\ n$, and $%
(n-1)(n-2)/2$, respectively) as follows:%
\[
\lbrack p_{a},p_{b}]=0,\hspace{10pt}  [x_{a},p_{b}]= h_{ab},\medskip 
\]%
\begin{equation}p^2[x_{a},x_{b}]=-S_{ab}-x_{a}p_{b}+x_{b}p_{a},\medskip  \label{2.12}
\end{equation}%
\[
\lbrack S_{ab},p_{c}]=0,\ \ \ \ [p^{2},S_{ab}]=0,
\]%
\[
p^{2}[S_{ab},x_{c}]=S_{ac}p_{b}-S_{bc}p_{a},\medskip 
\]%
\[
\lbrack S_{ab},S_{cd}]=h_{ac}S_{bd}-h_{bc}S_{ad}-h_{ad}S_{bc}+h_{bd}S_{ac}. 
\]%
\eject\noindent
Next, we pass to the Weyl algebra $\mathcal{D}_{e}(p^{2})$, which allows us
to define a full position operator:%
\begin{equation}
p^{2}y_{a}=p^{2}x_{a}-p_{a}(D-l)  \label{2.13}
\end{equation}%
such that $[y_{a},p_{b}]=g_{ab}.$ Here $l$ is a pure number to be determined
later.\medskip

This new operator can be regarded as unconstrained if we think of $D$ as
being defined in terms of $y_{a}$ by means of the relations $y\cdot
p=1-D+k+l $ (see equation (2.20) below). Then $y_{a}$ obeys the commutation relations:%
\begin{equation}
\lbrack p^{2},y_{a}]=-2p_{a},\ \ \ p^{2}[y_{a},y_{b}]=-S_{ab}.  \label{2.14}
\end{equation}%
\medskip \medskip Here as mentioned before, $D$ is the dilation operator
satisfying:
\begin{equation}
\lbrack D,p_{a}]=-p_{a},\ [D,x_{a}]=x_{a},\ [D,M_{ab}]=0.  \label{2.15}
\end{equation}%
(Additional commutation relations can be found in the appendix A.)

\bigskip \bigskip Consider now the full algebra $\mathcal{C}_{e}(p^{2}).$
Recall from equation (2.5) that the vector operator $q_{a}$ generates special
conformal transformations. Note that $p^{2}$ doesn't commute with $q_{a}$:%
\begin{equation}
\lbrack p^{2},q_{a}]=p_{a}(2D-1)-2p^{2}x_{a}\text{ \ \medskip }  \label{2.16}
\end{equation}%
\[ =p_{a}(2D-1)-(M_{ab}p^{b}+p^{b}M_{ab}).\]

Our immediate goal is to construct an operator $Q_{a}$ from $\mathcal{D}%
_{e}(p^{2})$ which obeys the same commutation relation with $p_{b} $ as does 
$q_{a},$ and therefore the difference $q_{a}-Q_{a}$ will then commute with $%
p_{a}$ and $p^{2}$.\medskip

\medskip Define:\smallskip 
\begin{equation}
Q_{a}=-y^{b}S_{ba}+\alpha \left( y\cdot y\right) p_{a}+\gamma \left( y\cdot
p\right) y_{a}  \label{2.17}
\end{equation}%
made from $M_{ab},\ D$ and $p_{a}$ only. When commuted with the $p$
operator, we obtain that:%
\[
\lbrack p_{a},Q_{b}]=S_{ab}-2\alpha y_{a}p_{b}-\gamma y_{b}p_{a}+\gamma
g_{ab}(1-y\cdot p). 
\]%
\pagebreak \bigskip \bigskip

Requiring that this commutation relation is the same as $[p_{a},q_{b}]$,\ we
obtain that

\begin{equation}
\alpha =-\frac{1}{2},\ \gamma =1,\ l= -k ,\text{ and }y\cdot p=1-D. 
\label{2.18}
\end{equation}%
\medskip Written in terms of the operator $x_{a}$, we have:%
\begin{equation}
Q_{a}=x^{b}S_{ab}-\frac{1}{2}\left( x\cdot x\right) p_{a}-x_{a}D+\frac{1}{2}%
p^{-2}p_{a}\left( D^{2}+D-k^{2}\right) ,  \label{2.19}
\end{equation}%
satisfying:%
\[
\lbrack p_{a},Q_{b}]=M_{ab}+Dg_{ab},\medskip 
\]%
\begin{equation}
\lbrack p^{2},Q_{a}]=p_{a}(2D-1)-2p^{2}x_{a},\medskip  \label{2.20}
\end{equation}%
\[
2p^{2}[Q_{a},Q_{b}]= - (n-3)(n-2) S_{ab}- 2S_{[b|d|}S^{cd}S_{a]c}. 
\]%

Now we define:%
\begin{equation}
R_{a}=q_{a}-Q_{a},  \label{2.21}
\end{equation}%
with the key commutation relations $[p_{a},R_{b}]=0$ and $[p^{2},R_{a}]=0$%
.\bigskip

We obtain in this way the algebra which we will call the $\mathcal{R}$%
-algebra. It is generated by $R_{a},\ p_{a}$, and $S_{ab}$ with the
commutation relations:%
\[
\lbrack p_{a},p_{b}]=0,\ \ \ [R_{a},p_{b}]=0,\ \ \ [S_{ab},p_{c}]=0,\medskip 
\]%
\[
p^{2}[S_{ab},R_{c}]=p^{2}h_{ac}R_{b}-p^{2}h_{bc}R_{a}-g_{ac}(R\cdot
p)p_{b}+g_{bc}(R\cdot p)p_{a},\medskip \medskip 
\]%
\begin{equation}
\lbrack S_{ab},S_{cd}]=h_{ac}S_{bd}-h_{bc}S_{ad}-h_{ad}S_{bc}+h_{bd}S_{ac}, 
\label{2.22}
\end{equation}%
\begin{eqnarray*}
2p^{2}[R_{a},R_{b}] &=&\left(\left( n-3\right) \left( n-2\right)  - 4\left( R\cdot p\right) \right)
S_{ab}+2S_{[b|d|}S^{cd}S_{a]c} \\
&&+ \left( R^{c}S_{ac}+S_{ac}R^{c}\right) p_{b}-\left(
R^{c}S_{bc}+S_{bc}R^{c}\right) p_{a}.
\end{eqnarray*}%
The last relation is derived by a very lengthy calculation.\bigskip
\bigskip
\eject\noindent
The following theorem is the main result of this work and
summarizes the results obtained in the style of [1].\bigskip

\noindent \textbf{THEOREM 2.1: }\textit{Define in}\texttt{\ }$\mathcal{C}%
_{e}(p^{2})$ \textit{the operator }$R_{a}$\textit{\ by the formula:}%
\begin{equation}
p^{2}R_{a}=p^{2}q_{a}-p^{2}x^{b}S_{ab}+\frac{1}{2}p^{2}\left( x\cdot
x\right) p_{a}+p^{2}x_{a}D-\frac{1}{2}p_{a}\left( D^{2}+D-k^{2}\right) . 
\nonumber
\end{equation}%
$R_{a}$ \textit{is translationally invariant:}%
\begin{equation}
\lbrack R_{a},p_{b}]=0.  \nonumber
\end{equation}%
\textit{Also,} $p^{4}(R_{a}-q_{a})\in \mathcal{D}_{e}$, \textit{and }$R_{a}$%
\textit{\ has the following commutation relations:}%
\begin{equation}
\lbrack D,R_{a}]=R_{a},\ \ \smallskip  \nonumber
\end{equation}%
\begin{equation}
\ p^{2}[S_{ab},R_{c}]=p^{2}h_{ac}R_{b}-p^{2}h_{bc}R_{a}-g_{ac}(R\cdot
p)p_{b}+g_{bc}(R\cdot p)p_{a},\smallskip   \nonumber
\end{equation}
\begin{eqnarray*}
2p^{2}[R_{a},R_{b}] &=&\left(\left( n-3\right) \left( n-2\right) - 4\left( R\cdot p\right)\right) 
S_{ab}+2S_{[b|d|}S^{cd}S_{a]c}  \\
&&+ \left( R^{c}S_{ac}+S_{ac}R^{c}\right) p_{b}-\left(
R^{c}S_{bc}+S_{bc}R^{c}\right) p_{a}.\nonumber
\end{eqnarray*}

\textit{The freedom in choosing }$R_{a}\in \mathcal{C}_{e}(p^{2})$, obeying%
\[
\lbrack p^{2},R_{a}]=0,\ \ p^{4}(R_{a}-q_{a})\in \mathcal{D}_{e}, 
\]%
is $R_{a}\rightarrow R_{a}+p^{4}g_{a}$ where $g_{a}\in \mathcal{P}_{e}$, and 
$[D,g_{a}]=5g_{a}$. \textit{Also, }$C_{e}(p^{2})$\textit{\ is generated by }$%
D$\textit{\ and the }$\mathcal{R}$\textit{-algebra.}

\subsection{Casimir Invariants}

We have now an algebra generated by the operators $R_{a},\ p_{a}$, and $%
S_{ab}$ that commute with the mass. In order to simplify the expressions, we
can introduce two new operators:%
\begin{equation}
S_{a}=R_{b}h_{\ a}^{b},  \label{2.26}
\end{equation}%
with $h_{ab}=g_{ab}-p^{-2}p_{a}p_{b}$, and%
\begin{equation}
S=R\cdot p=R_{a}p^{a}.  \label{2.27}
\end{equation}

Note that $p^{a}S_{a}=S_{a}p^{a}=0$ and $p^{a}S_{ab}=S_{ab}p^{a}=0.\bigskip $

The commutation relations become:%
\[
\lbrack S_{ab},S]=0,\ \ \ 2[S_{a},S]= 2S^{b}S_{ab}- (n -2)S_{a}, 
\]%
\begin{equation}
\lbrack S_{ab},S_{c}]=h_{ac}S_{b}-h_{bc}S_{a},\medskip  \label{2.28}
\end{equation}%
\[
\lbrack
S_{ab},S_{cd}]=h_{ac}S_{bd}-h_{bc}S_{ad}-h_{ad}S_{bc}+h_{bd}S_{ac}\medskip , 
\]%
\[
2p^{2}[S_{a},S_{b}]= ((n-3)\left( n-2\right)  - 4S) S_{ab} +2S_{[b|d|}S^{cd}S_{a]c} 
.\]%
\medskip

Note that by introducing these operators, $S_{a}$ transforms like a vector
with respect to the intrinsic spin $S_{ab}$, and the commutation relations are
significantly simpler. The generators of the $\mathcal{R}$-algebra are now the operators $p_a$, $%
S_{ab},\ S_{a},$ and $S $.  \medskip

A slight reformulation of this algebra can be achieved at the cost of introducing a scalar operator $m$ which is the square root of $p^2$, together with its inverse $m^{-1}$.  Then we may put $T_a = m^{-1} S_a$ and $s_a = m^{-1}p_a$.  The point here is that $S$, $S_{ab}$, $T_a$ and $s_a$ are dimensionless: $[D, S] = 0$, $[D, S_{ab}] = 0$, $ [D, T_a] = [D, s_a] =  0$. 
The commutation relations become:%
\begin{equation} [s_a, S] =0, \hspace{10pt} [s_a, s_b] = 0,   \hspace{10pt} [s_a, T_b] = 0, \hspace{10pt} [s_a, S_{bc}] = 0, \nonumber \end{equation}
\begin{equation} \nonumber \lbrack S_{ab}, S]=0,\ \ \ 2[T_{a},S]= 2T^{b}S_{ab}- (n-2)%
T_{a}, 
\end{equation}
\begin{equation}
\lbrack S_{ab}, T_{c}]=h_{ac}T_{b}-h_{bc}T_{a},\medskip  \label{2.28a}
\end{equation}%
\[
\lbrack
S_{ab}, S_{cd}]=h_{ac}S_{bd}-h_{bc}S_{ad}-h_{ad}S_{bc}+h_{bd}S_{ac}\medskip , 
\]%
\[ 2[T_{a}, T_{b}]=  ((n-3)\left( n-2\right) - 4S)
S_{ab}+2S_{[b|d|}S^{cd}S_{a]c},  
\]
\[ h_{ab} = g_{ab} - s_as_b, \hspace{10pt} s_as^a = 1, \hspace{10pt} h_{ab}s^b = 0, \hspace{10pt} S_{ab}s^b =0, \hspace{10pt} T_as^a =0.\nonumber \]
Note that the full conformal algebra is then recovered by adding in the two operators $D$ and $m$, which each commute with the operators $S$, $s_a$, $T_a$ and $S_{ab}$ and which obey the commutation relations:
\begin{equation}  [D, m] = - m.\label{2.28b}\end{equation}
In particular the momentum operator is recovered as $p_a = ms_a$.\eject\noindent
Finally note that, at least formally, if we write $m = e^x$ (which is intuitively reasonable when the mass is positive), then the commutation relation $[D, m] = - m$ is implied by the Heisenberg commutation relation:
\begin{equation} [x, D] = 1.\label{2.28c} \end{equation}
Thus we may consistently think of the enveloping algebra of the  conformal algebra as a straight-forward product of the Heisenberg algebra with the (dimensionless) algebra generated by the operators $S$, $s_a$, $T_a$ and $S_{bc}$.
\medskip

The following corollaries summarize our only general results on the Casimir
invariants.\medskip \bigskip

\textbf{Corollary 2.2: }\textit{The operator:}

\begin{equation}
C_{1}=S+\frac{1}{4}S_{cd}S^{cd},  \label{2.29}
\end{equation}%
\textit{is a Casimir invariant of the algebra }$C_{e}(p^{2}).\bigskip
\bigskip $

\textbf{Corollary 2.3: }$C_{1}$\textit{\ and }$p^{2}$\textit{\ are Casimir
invariants for the }$\mathcal{R}$\textit{-algebra.\bigskip \medskip }

$C_{1}$ can be used to reduce the number of generators just to the operators 
$S_{a}$ and $S_{ab}$, together with $C_1$ itself.\medskip \bigskip

It is known that $\mathbb{SO}(p+1,q+1)$ has rank $1+[\frac{n}{2}]$, this
being the number of Casimir invariants as well [3],\ [4]. In principle,
these invariants are given by the scalar operators formed from products of $%
M_{AB}$ with itself, but in practice this hasn't been very helpful in
finding the other Casimir invariants, due to the complexity of the
expressions involved.
Our attempts involved introducing yet another operator:%
\begin{equation}
U=S_{a}S^{a}.  \label{2.30}
\end{equation}%
\eject\noindent
The commutation relations involving the operator $U$ are as follows:%
\begin{equation}
\lbrack U,S_{ab}]=0,\ \ [U,S]=0, \end{equation}
\begin{equation} \label{2.31} 
p^{2}[U,S_{a}]=\left( \lambda \delta _{a}^{\ b}+\mu S_{a}^{\ b}+\nu S_{a}^{\
c}S_{c}^{\ b}+\rho S_{a}^{\ d}S_{d}^{\ c}S_{b}^{\ c}\right) S_{b},
\end{equation}%
where 
\begin{equation}
\begin{array}{l}
\lambda =\left( n-2\right) \left( n-1+2C_{1}-\frac{1}{2}S_{ab}S^{ab}\right)
\medskip , \\ 
\mu =n\left( n-1\right) +4C_{1}-S_{ab}S^{ab}\medskip , \\ 
\nu =3n-4\medskip , \\ 
\rho =2.%
\end{array}
\label{2.32}
\end{equation}

$U$ is a scalar with respect to $S_{ab}$ and $S$. It is expected that $U$
will be part of a Casimir invariant. To produce such an invariant, the
building blocks must be of the general form:%
\begin{equation}\label{2.33}
S^{a}F_{ab}^{\ }(S_{cd,}C_{1})S^{b}+G\left( S_{cd},C_{1}\right) , \end{equation}
where $F_{ab}^{\ }(S_{cd,}C_{1})$ is a function of $S_{cd}$ and $C_{1}$
only, with $a$ and $b$ free indices, e.g. $f(C_{1})S_{ad}S_{c}^{\
d}S_e^{\ c}S_{b}^{\ e},$ and $G\left( S_{cd},C_{1}\right) $ is a scalar function of $%
S_{cd}$ and $C_{1}$, e.g. $g(C_{1})S_{c}^{\ a}S_{b}^{\ c}S_{e}^{\ b}S_{a}^{\
e}.$\medskip

If necessary, one can assume that $F_{ab}^{\ }(S_{cd,}C_{1})$ is symmetric
in the indices $a$ and $b$, since the skew part allows $S^{a}FS^{b}$ to be
replaced by a commutator, which then can be assimilated into the $G$
term.  However except in dimensions $n  = p + q \le 5$, we have not yet been able to construct the remaining Casimirs.  The basic problem is the complexity of the relation (2.34).
\eject\noindent
\section{The Conformal Algebra in Three Dimensions\protect\bigskip}
\setcounter{equation}{0}
Consider the infinitesimal generators of the conformal group of a
three-dimensional space: the translations $p$, the special conformal
transformations $q$, the angular momentum $J$, and the dilation operator $D$%
. These generators form a ten-dimensional algebra. We first show a direct
construction of the $\mathcal{R}$-algebra and then we show that this
construction agrees with the specialization of the general case to
\thinspace $n=3$.\bigskip

\subsection{Basic Commutation Relations\protect\bigskip}

The basic commutation relations satisfied by these operators are:%
\begin{equation}
\begin{array}{l}
\lbrack D,p]=-p,\hspace{6pt}[D,J]=0,\hspace{6pt}[D,q]=q,\medskip \\ 
\lbrack J\cdot a,p\cdot b]=-(a\times b)\cdot p,\medskip \\ 
\lbrack J\cdot a,q\cdot b]=-(a\times b)\cdot q,\medskip \\ 
\lbrack J\cdot a,J\cdot b]=-(a\times b)\cdot J,\medskip \\ 
\lbrack p\cdot a,p\cdot b]=0,\hspace{6pt}[q\cdot a,q\cdot b]=0,\medskip \\ 
\lbrack p\cdot a,q\cdot b]=(a\cdot b)D-(a\times b)\cdot J.%
\end{array}
\label{3.1}
\end{equation}%
In this section we will use the notation $J\cdot a$ instead of $J_{a}$ and $%
(a\times b)\cdot J$ instead of $\varepsilon _{abc}J^{c}$ in order to
preserve some of the characteristics of a three-dimensional space.\medskip

These commutation relations ensure that the Jacobi identity is satisfied:%
\begin{equation}
0=[p\cdot a,[q\cdot b,q\cdot c]]+[q\cdot c,[p\cdot a,q\cdot b]]+[q\cdot
b,[q\cdot c,p\cdot a]].  \label{3.2}
\end{equation}

Introduce the following notation used for the remaining of this section:%
\begin{equation}
\lbrack p,\times q]=p\times q+q\times p,  \label{3.3}
\end{equation}%
for any two vectors $p,q.\medskip $

\subsection{Three-Dimensional Model\protect\bigskip}

We can construct a model in $\mathbb{R}^{3}$ that will satisfy the
commutation relation of the conformal algebra. This model has roots in
quantum mechanics, where $p$ is the momentum operator, $x$ is the position
operator, $J$ is the angular momentum and $D$ is the dilation operator:%
\begin{equation}
x_{a}=x_{a},\ p_{a}=\frac{\partial }{\partial x^{a}},\ \ J=x\times p,\
D=x\cdot p,  \label{3.4}
\end{equation}%
with the well-known commutation relations:\smallskip \smallskip 
\begin{equation}
\begin{array}{l}
\lbrack a\cdot J,p\cdot b]=-(a\times b)\cdot p,\medskip \\ 
\lbrack a\cdot J,x\cdot b]=-(a\times b)\cdot x,\medskip \\ 
\lbrack p_{a},x_{b}]=\delta _{ab}.%
\end{array}
\label{3.5}
\end{equation}%
\smallskip

Define now an operator $q$ that will satisfy the key relation of the conformal
algebra,  equation (3.1vi):%
\begin{equation}
q=\frac{1}{2}(x\times J+xD)  \label{3.6}
\end{equation}

Note that in this model we assumed that $J=x\times p,\ D=x\cdot p$, which
gives that $J\cdot p=p\cdot J=0$, and this would not generally be true. Once
we obtain an expression for the operator $q$ in terms of the generators of
the algebra, we will drop all assumptions on any particular model.\medskip

With these formulas, we obtain that the defining relation of the conformal
algebra is satisfied: 
\begin{equation}
\lbrack p\cdot a,q\cdot b]=(a\cdot b)D-(a\times b)\cdot J,  \label{3.7}
\end{equation}%
and we also obtain a candidate for the operator $q$: 
\begin{equation}
q=\frac{1}{2(p\cdot p)}\left[ (p\cdot J)J-p(J\cdot J)+pD(D+3)+2(p\times
J)(D+1)\right] ,  \label{3.8}
\end{equation}%
which can be seen not to commute with $p\cdot p$.\medskip

If we modify $q$ by a $p\cdot J$ term, we can define a new operator $Q$ by:
\smallskip 
\begin{equation}
2(p\cdot p)Q= 2(p\cdot J)J-p(J\cdot J)+pD(D+3)+2(p\times
J)(D+1) .  \label{3.9}
\end{equation}

This choice of the operator $Q$ still satisfies the conformal algebra:

\begin{equation}
\lbrack p\cdot a,Q\cdot b]=(a\cdot b)D-(a\times b)\cdot J,  \label{3.10}
\end{equation}%
and%
\begin{equation}
\lbrack p\cdot p,Q]=p(2D+1)+2(p\times J)=[p\cdot p,q].  \label{3.11}
\end{equation}

Since neither $q$ nor $Q$ commute with $p\cdot p$, this leads us to
introducing a new operator, $R$.

\subsection{The $\mathcal{R}$-algebra}

Construct the following operator: $R=q-Q$, with the commutator:

\begin{equation}
\lbrack p\cdot a,R\cdot b]=0.  \label{3.12}
\end{equation}

$R$ is a vector operator, so we have the commutation relation: 
\begin{equation}
\lbrack J\cdot a,R\cdot b]=-(a\times b)\cdot R.  \label{3.13}
\end{equation}

Moreover, $R$ has the same dimension as does $q$, so we have: 
\begin{equation}
\lbrack D,R]=R.  \label{3.14}
\end{equation}%
\medskip

\noindent After lengthy calculations, we obtain: 
\begin{equation}
(p\cdot p)^{2}(Q\times Q)=2(p\cdot J)p,  \label{3.15}
\end{equation}

This result can be used to write: 
\begin{equation}
R\times R=(q-Q)\times (q-Q)=Q\times Q-(q\times Q+Q\times q),  \label{3.16}
\end{equation}%
where we took into account that $q\times q=0$, and:%
\begin{equation}
(p\cdot p)^{2}(R\times R)=2(p\cdot p)(p\cdot J)R-(p\cdot p)(p\times
R)-2(p\cdot J)p.  \label{3.17}
\end{equation}

\begin{center}
\bigskip
\end{center}

\subsection{Casimir Operators}

We have constructed the subalgebra with generators $p,\ J,$ and $R$, and
obtained the complete commutation relations. This algebra, as desired, has
the mass of the system as one of the Casimir operators.\medskip

For the remaining of this section, we are trying to construct the Casimir
invariants of the algebra. The following commutation relations have been
established:%
\[
\lbrack R\cdot p,p]=0,\ \ \ [R\cdot p,J]=0\medskip , 
\]%
\begin{equation}
(p\cdot p)[R\cdot p,R]=-2(p\cdot J)(p\times R)+(R\cdot p)p-(p\cdot
p)R\medskip ,  \label{3.18}
\end{equation}%
\[
\lbrack R\cdot R,p]=0,\ \ \ [R\cdot R,J]=0\medskip , 
\]%
\[
(p\cdot p)^{2}[R\cdot R,R]=-2\left( p\cdot J\right) \left( p\times R\right)
-(p\cdot p)R+(R\cdot p)p. 
\]

Since not all these commutators vanish, we introduce a new operator:%
\begin{equation}
S=R-\lambda \frac{p}{p\cdot p}  \label{3.19}
\end{equation}%
with%
\begin{equation}
S\cdot p=R\cdot p-\lambda ,  \label{3.20}
\end{equation}%
\medskip 
\[
S\cdot S=R\cdot R-2\lambda (p\cdot p)^{-1}(R\cdot p)+(p\cdot p)^{-1}\lambda
^{2}. 
\]

By commuting it with the new generators of the algebra, namely $R,\ S$, and $%
p$, we obtain that $S\cdot S$ is a Casimir operator if $\lambda =1/2.$ Note
that the algebra is written now in terms of $S,\ p$, and $J$.\bigskip

With $\lambda $ fixed at $1/2$, we can show that the two Casimir operators
of the algebra $\mathcal{C}_{e}$ have the simple expressions: 
\begin{equation}
\begin{array}{l}
C_{1}=S\cdot S\medskip , \\ 
C_{2}=S\cdot p+\frac{1}{p\cdot p}(p\cdot J)^{2}.%
\end{array}
\label{3.21}
\end{equation}%
The Casimir invariants of the $\mathcal{R}$-algebra (more appropriately now
called $\mathcal{S}$-algebra) are $C_{1},\ C_{2}$ and $p\cdot p$.

\subsection{Applying the $n$-d Theory to the 3-d Theory\protect\bigskip}

As mentioned before, the two theories have been developed independently. In
this subsection we show that the two approaches are consistent with each
other. The main test is showing that the operators $Q$ and $R\,\ $satisfy
the same relations in both theories, and that the Casimir invariant in the $%
n $-dimensional theory when applied to $n=3$ matches the one derived in the
original three-dimensional theory.\medskip \medskip

We consider the following relation that will make the connection between the
two cases:%
\begin{equation}
M_{ab}=-\epsilon _{abc}J^{c}\ \ \ \text{or \ \ }J_{c}=-\frac{1}{2}%
\varepsilon _{abc}M^{ab}.  \label{3.22}
\end{equation}%
\medskip

One can show fairly easily that the commutation relations (3.1) are
satisfied by these expressions.\medskip

We also have:%
\[
p^{2}x=-(p\times J)-p\medskip , 
\]%
\begin{equation}
p^{2}L_{ab}=\varepsilon _{abc}(p\cdot J)p^{c}-p^{2}\varepsilon
_{abc}J^{c}=(p\cdot J)\left( a\times b\right) \cdot p-(p\cdot p)(a\times
b)\cdot J\medskip ,  \label{3.23}
\end{equation}%
\[
p^{2}S_{ab}=-\varepsilon _{abc}(p\cdot J)p^{c}=-(p\cdot J)(a\times b)\cdot
p, 
\]%
which yields:%
\begin{equation}
S_{ab}S^{ab}=2p^{-2}(p\cdot J)^{2}.  \label{3.24}
\end{equation}

Using these formulas in equation (2.19), we obtain the following expression for the $%
Q $ operator:%
\begin{eqnarray}
2(p\cdot p)Q &=&2(p\cdot J)J-p(J\cdot J)+pD(D+3)  \label{3.25} \\
&&+2(p\times J)(D+1)+2p-(p\cdot p)^{-1}(p\cdot J)^{2}p.  \nonumber
\end{eqnarray}

Note that the difference between equations (3.25) and (2.19) is given by the last two
terms, $2p-(p\cdot p)^{-1}(p\cdot J)^{2}p$; as long as they commute with $p$
(and $p^{2}$), these terms will not change what we require from the $R$
operator (that is to commute with $p^{2}$). The choice of the $q$ and $Q~$%
operators (and, therefore, the Casimir operators), is not unique.\medskip
\bigskip

In the $n$-dimensional case, we obtained that one of the Casimir operators
was: 
\begin{equation}
C=R\cdot p+\frac{1}{4}S_{ab}S^{ab}. \label{3.25a} 
\end{equation}

The corresponding Casimir operator in three-dimensions has been obtained
from this general one and has the form:%
\begin{equation}
C=R\cdot p+\frac{1}{2}S_{ab}S^{ab}. \label{3.25b}
\end{equation}

Although these two relations seem slightly different, they can be shown to
be the same. If we prime the quantities arising from letting $n=3$ in the
general theory, we have that:%
\begin{equation}
2p^{2}R^{\prime }=2p^{2}R+p^{-2}(p\cdot J)^{2}p-2p,  \label{3.26}
\end{equation}%
and thus:%
\begin{equation}
2p^{2}(R^{\prime }\cdot p)=2p^{2}(R\cdot p)+(p\cdot J)^{2}-2p^{2}  \label{3.27}
\end{equation}%
or%
\begin{eqnarray}
R^{\prime }\cdot p &=&R\cdot p+\frac{1}{2}(p\cdot J)^{2}-1  \label{3.28} \\
&=&R\cdot p+\frac{1}{4}S_{ab}S^{ab}-1  \nonumber
\end{eqnarray}%
which means that%
\begin{equation}
R^{\prime }\cdot p+\frac{1}{4}S_{ab}S^{ab}=R\cdot p+\frac{1}{2}S_{ab}S^{ab},
\label{3.29}
\end{equation}%
(where we ignored the constant $-1$).\medskip

This shows that the two Casimirs operators are the same, hence the two
theories agree.

\begin{center}
\bigskip \pagebreak
\end{center}

\section{The Conformal Algebra in Two Dimensions}
\setcounter{equation}{0}
Consider the case, $n=p+q=2$. This situation is relevant for the Lorentz
group $\mathbb{SO}(1,3)$ (or $\mathbb{SO}(3,1)$) and for the ultra-hyperbolic group $\mathbb{SO}(2,2)$
which is used in studying solitons and integrable systems [12], [13]. The results are obtained by
directly applying the general theory to $n=2$; as a consequence, we only
describe how the quantities considered there change and their new
commutation relations.\medskip

In two dimensions, any skew quantity with two indices must be a multiple of
the completely antisymmetric tensor $\varepsilon _{ab}$. We have thus:%
\begin{equation}
M_{ab}=J\varepsilon _{ab}.  \label{4.1}
\end{equation}

The $\varepsilon _{ab}$ tensor satisfies the following identity:%
\[
\varepsilon _{ab}\varepsilon ^{cd}=\sigma \left( \delta _{a}^{c}\delta
_{b}^{d}-\delta _{a}^{d}\delta _{b}^{c}\right) , 
\]%
with $\sigma =-1$ for $(p,q)=(1,1)$, and $\sigma =1$ for $(p,q)=(0,2)$ or $%
(2,0).$ In this section we consider $\sigma =1$.\medskip

We have the following basic commutation relations for the $\mathcal{C}_{e}$
algebra:%
\begin{equation}
\begin{array}{l}
\lbrack p_{a},p_{b}]=0,\ \ [q_{a},q_{b}]=0,\medskip \\ 
\lbrack J,p_{a}]=\varepsilon _{a}^{\ b}p_{b},\ \ [J,q_{a}]=\varepsilon
_{a}^{\ b}q_{b},\medskip \\ 
\lbrack D,p_{a}]=-p_{a},\ \ [D,q_{a}]=q_{a},\medskip \\ 
\lbrack J,J]=0,\ \ [D,D]=0,\ [D,J]=0,\medskip \\ 
\lbrack p_{a},q_{b}]=\varepsilon _{ab}J+g_{ab}D.%
\end{array}
\label{4.2}
\end{equation}

The new operators corresponding to (2.9)-(2.11) are now:%
\begin{equation}
\begin{array}{l}
p^{2}x_{a}=\varepsilon _{ab}Jp^{b}+\frac{1}{2}p_{a},\medskip \\ 
L_{ab}=J\varepsilon _{ab}=M_{ab},\medskip \\ 
S_{ab}=0.%
\end{array}
\label{4.3}
\end{equation}

The operator $Q_{a}$ defined in (2.19) becomes in this case:%
\begin{equation}
2p^{2}Q_{a}=-J^{2}p_{a}+p_{a}D^{2}-2\varepsilon _{ab}Jp^{b}D,  \label{4.4}
\end{equation}%
which can be shown to satisfy:%
\begin{equation}
\lbrack p_{a},Q_{b}]=\varepsilon _{ab}J+g_{ab}D,  \label{4.5}
\end{equation}%
as desired.\medskip

This new operator also satisfies:%
\begin{equation}
\lbrack Q_{a},Q_{b}]=0,  \label{4.6}
\end{equation}%
which is consistent with equation (2.22iii) since $S_{ab}=0$.\medskip

As in the previous cases, let $R_{a}=q_{a}-Q_{a}$. The $\mathcal{R}$-algebra
will be now generated by the operators $p_{a},\ R_{a}$ and $J$ satisfying
the following commutation relations:%
\begin{equation}
\begin{array}{l}
\lbrack R_{a},p_{b}]=0,\medskip \\ 
\lbrack R_{a},R_{b}]=0,\medskip \\ 
\lbrack R_{a},J]=0,\medskip \\ 
\lbrack p_{a},p_{b}]=0,\medskip \\ 
\lbrack J,p_{a}]=\varepsilon _{a}^{\ b}p_{b}.%
\end{array}
\label{4.7}
\end{equation}

It is easy to show that in this case the two Casimir invariants of the $%
\mathcal{C}_{e}(p^{2})$ algebra are:%
\begin{equation}
\begin{array}{l}
C_{1}=R\cdot p, \\ 
C_{2}=R\cdot R,%
\end{array}
\label{4.8}
\end{equation}%
and the Casimir invariants of the $\mathcal{R}$-algebra are $C_{1},\ C_{2}$,
and $p^{2}$.

\begin{center}
\pagebreak
\end{center}

\section{Conclusions}

In this work we have given a complete description of the $\mathcal{R}$%
-algebras in $n$-dimensions and the lower-dimensional cases: $n=2,$ and$\
n=3.\medskip $

In reference [1], the $\mathcal{R}$-algebra, for the case $n=4$, was used to
construct the family of unitary irreducible representations that constitute
the discrete series of the conformal group. We expect that in general the $%
\mathcal{R}$-algebras constructed here can be used in the same way.\smallskip

Finding the Casimir operators for these theories was not a trivial task, but
it proved to be extremely difficult in the general case, preventing us from
obtaining a complete result.

\smallskip We have studied the 5-dimensional case as well; these results
will be presented in a separate paper, as the approach is quite different,
requiring spinor techniques to make the calculations tractable. Such
techniques can also be used to simplify the original 4-dimensional
case.\medskip

We are currently investigating the possibility of extending these results to
super-conformal algebras as well [9], [10].

\pagebreak

\pagebreak

\begin{center}
{\Large Appendix A: }{\Large $n$-dimensional case}
\end{center}
\renewcommand{\theequation}{A.\arabic{equation}}
\setcounter{equation}{0}
\begin{equation}
\lbrack p_{a},p_{b}]=0,\ \ [q_{a},q_{b}]=0,  \label{A.1}
\end{equation}%
\begin{equation}
\lbrack M_{ab},p_{c}]=g_{ac}p_{b}-g_{bc}p_{a},  \label{A.2}
\end{equation}%
\begin{equation}
\lbrack M_{ab},q_{c}]=g_{ac}q_{b}-g_{bc}q_{a},  \label{A.3}
\end{equation}%
\begin{equation}
\lbrack M_{ab},M_{cd}]=g_{ac}M_{bd}-g_{bc}M_{ad}-g_{ad}M_{bc}+g_{bd}M_{ac}, 
\label{A.4}
\end{equation}%
\begin{equation}
\lbrack D,p_{a}]=-p_{a},\ \ \ \ \ [D,q_{a}]=q_{a},  \label{A.5}
\end{equation}%
\begin{equation}
\lbrack D,D]=0,\ \ \ \ \ \ \ [D,M_{ab}]=0,  \label{A.6}
\end{equation}%
\begin{equation}
\lbrack p_{a},q_{b}]=M_{ab}+g_{ab}D,  \label{A.7}
\end{equation}%
\begin{equation}
M_{ab}p^{a}p^{b}=0,  \label{A.8}
\end{equation}%
\begin{equation}
p^{2}x_{a}=\frac{1}{2}\left( M_{ab}p^{b}+p^{b}M_{ab}\right)
=M_{ab}p^{b}+kp_{a},  \label{A.9}
\end{equation}%
\begin{equation}
\lbrack p^{2},q_{a}]=p_{a}(2D-1)-2p^{2}x_{a},  \label{A.10}
\end{equation}%
\begin{equation}
\lbrack M_{ab},x_{c}]=g_{ac}x_{b}-g_{bc}x_{a},  \label{A.11}
\end{equation}%
\begin{equation}
p^{2}[x_{a},x_{b}]=-M_{ab},  \label{A.12}
\end{equation}%
\begin{equation}
p^{4}(x\cdot x)=(M_{ab}p^{b})(M_{\ c}^{a}p^{c})-\frac{(n-1)^{2}}{4}p^{2}, 
\label{A.13}
\end{equation}%
\begin{equation}
p^{2}[x\cdot x,x_{a}]=2x^{b}M_{ab}-2kx_{a},  \label{A.14}
\end{equation}%
\begin{equation}
\left[ x\cdot x,p_{a}\right] =2x_{a},  \label{A.15}
\end{equation}%
\begin{equation}
\lbrack p^{2},x_{a}]=0,  \label{A.16}
\end{equation}%
\begin{equation}
[x_{a}, p_{b}] = g_{ab}-p^{-2} p_{a}p_{b} := h_{ab},  \label{A.17}
\end{equation}%
\begin{equation}
x\cdot p=-p\cdot x:=k=\frac{n-1}{2},  \label{A.18}
\end{equation}%
\begin{equation}
L_{ab}=x_{a}p_{b}-x_{b}p_{a},\ \ \ L_{ab}p^{a}p^{b}=0,  \label{A.19}
\end{equation}%
\begin{equation}
\lbrack M_{ab},L_{cd}]=g_{ac}L_{bd}-g_{bc}L_{ad}-g_{ad}L_{bc}+g_{bd}L_{ac}, 
\label{A.20}
\end{equation}%
\begin{equation}
\lbrack L_{ab},p_{c}]=g_{ac}p_{b}-g_{bc}p_{a},  \label{A.21}
\end{equation}%
\begin{equation}
S_{ab}=M_{ab}-L_{ab},  \label{A.22}
\end{equation}%
\begin{equation}
\lbrack M_{ab},S_{cd}]=g_{ac}S_{bd}-g_{bc}S_{ad}-g_{ad}S_{bc}+g_{bd}S_{ac}, 
\label{A.23}
\end{equation}%
\begin{equation}
\lbrack L_{ab},x_{c}]=g_{ac}x_{b}-g_{bc}x_{a}-p^{-2}\left(
S_{ac}p_{b}-S_{bc}p_{a}\right) ,  \label{A.24}
\end{equation}%
\begin{equation}
\lbrack S_{ab},p_{c}]=0,\ \ \ \ S_{ab}p^{b}=0,\ \ [p^{2},S_{ab}]=0, 
\label{A.25}
\end{equation}%
\begin{equation}
p^{2}[S_{ab},x_{c}]=S_{ac}p_{b}-S_{bc}p_{a},  \label{A.26}
\end{equation}%
\begin{equation}
\lbrack S_{ab},S_{cd}]=h_{ac}S_{bd}-h_{bc}S_{ad}-h_{ad}S_{bc}+h_{bd}S_{ac}, 
\label{A.27}
\end{equation}%
\begin{equation}
\lbrack S_{ab},x^{b}]=0,  \label{A.28}
\end{equation}%
\begin{equation}
p^{2}y_{a}=p^{2}x_{a}-p_{a}D,  \label{A.29}
\end{equation}%
\begin{equation}
\lbrack p^{2},y_{a}]=-2p_{a},\ \ p^{2}[y_{a},y_{b}]=-S_{ab},  \label{A.30}
\end{equation}%
\begin{equation}
y_{a}p_{b}-y_{b}p_{a}=L_{ab},  \label{A.31}
\end{equation}%
\begin{equation}
\lbrack D,M_{ab}]=0=[D,S_{ab}]=[D,L_{ab}],  \label{A.32}
\end{equation}%
\begin{equation}
\lbrack D,p^{2}]=-2p^{2},\ [D,p^{-2}]=2p^{-2},  \label{A.33}
\end{equation}%
\begin{equation}
\lbrack D^{2},p^{2}]=4p^{2}(1-D),  \label{A.34}
\end{equation}%
\begin{equation}
\lbrack D,p^{-2}p_{a}]=p^{-2}p_{a},  \label{A.35}
\end{equation}%
\begin{equation}
\lbrack D^{2},p_{a}]=p_{a}(-2D+1),  \label{A.36}
\end{equation}%
\begin{equation}
\lbrack D,y_{a}]=y_{a},\ \ \ \ \ [D,x_{a}]=x_{a},\ \ \ [D,Q_{a}]=Q_{a}, 
\label{A.37}
\end{equation}%
\begin{equation}
\lbrack D,x\cdot x]=2\left( x\cdot x\right) ,  \label{A.38}
\end{equation}%
\begin{equation}
f(D)x_{a}=x_{a}f(D+1),\ f(D)p_{a}=p_{a}f(D-1),  \label{A.39}
\end{equation}%
for any polynomial function $f$. 
\begin{equation}
Q_{a}=x^{b}S_{ab}-\frac{1}{2}\left( x\cdot x\right) p_{a}-x_{a}D+\frac{1}{2}%
p^{-2}p_{a}\left( D^{2}+D-k^{2}\right) ,  \label{A.40}
\end{equation}%
\begin{equation}
\lbrack p^{2},Q_{a}]=[p^{2},q_{a}],\ \ \ \ [p_{a},Q_{b}]=[p_{a},q_{b}], 
\label{A.41}
\end{equation}%
\begin{equation}
\lbrack R_{a},p_{b}]=0,  \label{A.42}
\end{equation}%
\begin{equation}
\lbrack D,R_{a}]=0,  \label{A.43}
\end{equation}%
\begin{equation}
\lbrack R_{a},p^{2}]=0,  \label{A.44}
\end{equation}%
\begin{equation}
\lbrack M_{ab},R_{c}]=g_{ac}R_{b}-g_{bc}R_{a},  \label{A.45}
\end{equation}%
\begin{equation}
p^{2}[x_{a},R_{b}]=g_{ab}(R\cdot p)-R_{a}p_{b},  \label{A.46}
\end{equation}%
\begin{equation}
p^{2}[S_{ab},R_{c}]=p^{2}h_{ac}R_{b}-p^{2}h_{bc}R_{a}-g_{ac}(R\cdot
p)p_{b}+g_{bc}(R\cdot p)p_{a},  \label{A.47}
\end{equation}%
\vfill\eject
\begin{equation}\nonumber
2p^{2}[R_{a},R_{b}]= ((n-3)( n-2) - 4R\cdot p)
S_{ab}+ 2S_{[b|d|}S^{cd}S_{a]c} \end{equation}
\begin{equation} + \left(
R^{c}S_{ac}+S_{ac}R^{c}\right) p_{b}-\left( R^{c}S_{bc}+S_{bc}R^{c}\right)
p_{a},%
\label{A.48}
\end{equation}%
\begin{equation}
\lbrack S_{a}^{\ c},S_{bc}]=(n-3)S_{ab},  \label{A.49}
\end{equation}%
\begin{equation}
\lbrack S_{ab},x\cdot x]=2x^{c}p_{b}S_{ac}-2x^{c}p_{a}S_{bc}-2S_{ab}, 
\label{A.50}
\end{equation}%
\begin{equation}
R\cdot x=x\cdot R-p^{-2}(n-1)(R\cdot p),  \label{A.51}
\end{equation}%
\begin{equation}
\lbrack x_{a},R\cdot p]=0,  \label{A.52}
\end{equation}%
\begin{equation}
\lbrack x\cdot R,p_{a}]=R_{a}-p^{-2}p_{a}(R\cdot p),  \label{A.53}
\end{equation}%
\begin{equation}
\lbrack S_{ac},R^{c}]=\left( n-2\right) [p^{-2}(R\cdot p)p_{a}-R_{a}], 
\label{A.54}
\end{equation}%
\begin{equation}
\lbrack R\cdot p,R_{a}]=-\frac{1}{2}\left( R^{b}S_{ab}+S_{ab}R^{b}\right) , 
\label{A.55}
\end{equation}%
\begin{equation}
\lbrack S_{ab},R\cdot p]=0,  \label{A.56}
\end{equation}%
\begin{equation}
\left[ S_{ac}S^{ac},R_{b}\right] =2\left( S_{bc}R^{c}+R^{c}S_{bc}\right) , 
\label{A.57}
\end{equation}%
\begin{equation}
\lbrack S_{ab}S^{ab},S_{cd}]=0,  \label{A.58}
\end{equation}%
\begin{equation}
S_{a}=R_{b}h_{\ a}^{b},\ \ S_{a}p^{a}=0,  \label{A.59}
\end{equation}%
\begin{equation}
S=R\cdot p,\ \ [S,S]=0,  \label{A.60}
\end{equation}%
\begin{equation}
\lbrack S_{ab},S]=0,\ \ [S_{a},S]=S^{b}S_{ab}-\frac{n-2}{2}S_{a},  \label{A.61}
\end{equation}%
\begin{equation}
\lbrack S_{ab},S_{c}]=h_{ac}S_{b}-h_{bc}S_{a},  \label{A.62}
\end{equation}%
\begin{equation}
p^{2}[S_{a},S_{b}]=\frac{(n-3)\left( n-2\right) }{2}%
S_{ab}+S_{[b|d|}S^{cd}S_{a]c}-2SS_{ab},  \label{A.63}
\end{equation}%
\begin{equation}
U=S_{a}S^{a},  \label{A.64}
\end{equation}%
\begin{equation}
\lbrack U,S_{ab}]=0=[U,S]=0,  \label{A.65}
\end{equation}%
\begin{equation}
p^{2}[U,S_{a}] =    2S_{a}^{\ d}S_{d}^{\ c}S_{c}^{\ b}S_b \nonumber +\left( n\left( n-1\right) +4C_{1}-S_{cd}S^{cd}\right) S_{a}^{\
b}S_b \end{equation} \begin{equation} +(3n-4)S_{a}^{\ c}S_{c}^{\ b}S_b +  \left( n-2\right) \left( n-1+2C_{1}-\frac{1}{2}%
S_{cd}S^{cd}\right) S_{a}  
\label{A.66} 
\end{equation}
\bigskip

\begin{center}
{\Large Appendix B: 3-dimensional case}\bigskip
\renewcommand{\theequation}{B.\arabic{equation}}
\setcounter{equation}{0}
\begin{equation}
\lbrack D,p]=-p,\ \ \ [D,J]=0,\ \ \ [D,q]=q,  \label{B.1}
\end{equation}%
\begin{equation}
\lbrack J\cdot a,p\cdot b]=-(a\times b)\cdot p,  \label{B.2}
\end{equation}%
\begin{equation}
\lbrack J\cdot a,q\cdot b]=-(a\times b)\cdot q,  \label{B.3}
\end{equation}%
\begin{equation}
\lbrack J\cdot a,J\cdot b]=-(a\times b)\cdot J,  \label{B.4}
\end{equation}%
\begin{equation}
\lbrack p\cdot a,p\cdot b]=0,\hspace{6pt}[q\cdot a,q\cdot b]=0,  \label{B.5}
\end{equation}%
\begin{equation}
\lbrack p\cdot a,q\cdot b]=(a\cdot b)D-(a\times b)\cdot J,  \label{B.6}
\end{equation}%
\begin{equation}
\lbrack p,\times q]:=p\times q+q\times p,  \label{B.7}
\end{equation}%
\begin{equation}
(a\times b)\cdot (J\times p+p\times J)=-2(a\times b)\cdot p,  \label{B.8}
\end{equation}%
\begin{equation}
\lbrack J,\times p]=-2p,\ \ [J,\times q]=-2q,  \label{B.9}
\end{equation}%
\begin{equation}
J\times J=-J,  \label{B.10}
\end{equation}%
\begin{equation}
\lbrack J,p\cdot p]=0,  \label{B.11}
\end{equation}%
\begin{equation}
\lbrack p,J\cdot J]=p\times J-J\times p=2p\times J+2p,  \label{B.12}
\end{equation}%
\begin{equation}
\lbrack q,J\cdot J]=q\times J-J\times q=2q\times J+2q,  \label{B.13}
\end{equation}%
\begin{equation}
\lbrack J,J\cdot J]=0,  \label{B.14}
\end{equation}%
\begin{equation}
\lbrack p\cdot p,q]=pD+Dp+p\times J-J\times p=p(2D+1)+2p\times J,  \label{B.15}
\end{equation}%
\begin{equation}
2(p\cdot p)Q=2(p\cdot J)J-p(J\cdot J)+pD(D+3)+2(p\times J)(D+1),  \label{B.16}
\end{equation}%
\begin{equation}
\lbrack p\cdot a,Q\cdot b]=(a\cdot b)D-(a\times b)\cdot J,  \label{B.17}
\end{equation}%
\begin{equation}
\lbrack p\cdot p,Q]=p(2D+1)+2(p\times J)=[p\cdot p,q],  \label{B.18}
\end{equation}%
\begin{equation}
J\cdot p=p\cdot J,\hspace{6pt}J\cdot q=q\cdot J,  \label{B.19}
\end{equation}%
\begin{equation}
\lbrack J\cdot p,p]=0,[J\cdot p,J]=0,[J\cdot p,q]=J(D-1)-q\times p, 
\label{B.20}
\end{equation}%
\begin{equation}
\lbrack q,\times p]=[p,\times q]=-2J,  \label{B.21}
\end{equation}%
\begin{equation}
\left( p\times J\right) \times J=-p\times J+(p\cdot J)J-p(J\cdot J), 
\label{B.22}
\end{equation}%
\begin{equation}
J\times (p\times J)=-p\times J-(p\cdot J)J+p(J\cdot J),  \label{B.23}
\end{equation}%
\begin{equation}
p\times (p\times J)=(p\cdot J)p-(p\cdot p)J,  \label{B.24}
\end{equation}%
\begin{equation}
(p\times J)\times p=-(p\cdot J)p+(p\cdot p)J,  \label{B.25}
\end{equation}%
\begin{equation}
\lbrack p\cdot J,J\cdot J]=0,  \label{B.26}
\end{equation}%
\begin{equation}
p\cdot (p\times J)=0,  \label{B.27}
\end{equation}%
\begin{equation}
(p\times J)\cdot p=-2p^{2},  \label{B.28}
\end{equation}%
\begin{equation}
(p\times J)\cdot J=-p\cdot J,  \label{B.29}
\end{equation}%
\begin{equation}
J\cdot (p\times J)=-p\cdot J,  \label{B.30}
\end{equation}%
\begin{equation}
(p\times J)\times (p\times J)=(p\cdot p)J,  \label{B.31}
\end{equation}%
\begin{equation}
(p\times J)\cdot (p\times J)=(p\cdot p)(J\cdot J)-(p\cdot J)^{2},  \label{B.32}
\end{equation}%
\begin{equation}
\lbrack p\cdot p,J\cdot J]=0,  \label{B.33}
\end{equation}%
\begin{equation}
x_{a}p_{b}-x_{b}p_{a}=(a\times b)\cdot (x\times p),  \label{B.34}
\end{equation}%
\begin{equation}
\left[ p_{a},(p\times J)_{b}\right] =(a\cdot b)(p\cdot p)-p_{a}p_{b}, 
\label{B.35}
\end{equation}%
\begin{equation}
\lbrack J\cdot q,p]=-(p\times q)-J(D+1),  \label{B.36}
\end{equation}%
\begin{equation}
(p\times J)\times q=(p\cdot q)J-(q\cdot J)p-2(p\times q)-J(D+1),  \label{B.37}
\end{equation}%
\begin{equation}
p\times (q\times J)=(p\cdot J)q-(p\times q)-(p\cdot q)J,  \label{B.38}
\end{equation}%
\begin{equation}
(p\times q)\times J=(p\cdot J)q-(q\cdot J)p-2(p\times q)-J(D+1),  \label{B.39}
\end{equation}%
\begin{equation}
\lbrack J\cdot q,p]=-(p\times q)-J(D+1),  \label{B.40}
\end{equation}%
\begin{equation}
(p\cdot p)^{2}(Q\times Q)=2(p\cdot J)p,  \label{B.41}
\end{equation}%
\begin{equation}
R=q-Q,  \label{B.42}
\end{equation}%
\begin{equation}
\lbrack p\cdot p,R]=0,\ [p\cdot a,R\cdot b]=0,  \label{B.43}
\end{equation}%
\begin{equation}
\lbrack D,R]=R,  \label{B.44}
\end{equation}%
\begin{equation}
\lbrack J\cdot a,R\cdot b]=-(a\times b)\cdot R,  \label{B.45}
\end{equation}%
\begin{equation}
(p\cdot p)^{2}(R\times R)=2(p\cdot p)(p\cdot J)R-(p\cdot p)(p\times
R)-2(p\cdot p)(p\cdot J)p,  \label{B.46}
\end{equation}%
\begin{equation}
(p\times R)\times p=(p\cdot p)R-(p\cdot R)p,  \label{B.47}
\end{equation}%
\begin{equation}
p\times (p\times R)=(p\cdot R)p-(p\cdot p)R,  \label{B.48}
\end{equation}%
\begin{equation}
\lbrack p\cdot J,R]=p\times R,  \label{B.49}
\end{equation}%
\begin{equation}
\lbrack (p\cdot J)^{2},R]=2(p\cdot J)(p\times R)+(p\cdot p)R-(R\cdot p)p, 
\label{B.50}
\end{equation}%
\begin{equation}
(p\times R)\times R=R(p\cdot R)-(R\cdot R)p,  \label{B.51}
\end{equation}%
\begin{equation}
R\times (p\times R)=(R\cdot R)p-(p\cdot R)R,  \label{B.52}
\end{equation}%
\begin{equation}
\lbrack R\cdot p,p]=0,\ \ \ [R\cdot p,J]=0,  \label{B.53}
\end{equation}%
\begin{equation}
(p\cdot p)[R\cdot p,R]=-2(p\cdot J)(p\times R)+(R\cdot p)p-(p\cdot p)R, 
\label{B.54}
\end{equation}%
\begin{equation}
\lbrack R\cdot R,p]=0,\ \ \ [R\cdot R,J]=0,  \label{B.55}
\end{equation}%
\begin{equation}
S=R-\frac{1}{2}\frac{p}{p\cdot p},  \label{B.56}
\end{equation}%
\begin{equation}
S\cdot p=R\cdot p-\frac{1}{2},  \label{B.57}
\end{equation}%
\begin{equation}
(p\cdot p)(S\cdot S)=(p\cdot p)(R\cdot R)-(R\cdot p)+\frac{1}{4}.  \label{B.58}
\end{equation}%
\bigskip
\eject\noindent
{\Large Appendix C: 2-dimensional case}
\renewcommand{\theequation}{C.\arabic{equation}}
\setcounter{equation}{0}
\begin{equation}
M_{ab}=J\varepsilon _{ab},  \label{C.1}
\end{equation}%
\begin{equation}
\lbrack p_{a},p_{b}]=0,\ \ \ \ \ [q_{a},q_{b}]=0,  \label{C.2}
\end{equation}%
\begin{equation}
\lbrack J,p_{c}]=\varepsilon _{c}^{\ b}p_{b},\ [J,q_{c}]=\varepsilon _{c}^{\
b}q_{b},  \label{C.3}
\end{equation}%
\begin{equation}
\lbrack D,p_{a}]=-p_{a},\ \ \ \ [D,q_{a}]=q_{a},  \label{C.4}
\end{equation}%
\begin{equation}
\lbrack J,J]=0=[D,D]=[D,J],  \label{C.5}
\end{equation}%
\begin{equation}
\lbrack p_{a},q_{b}]=\varepsilon _{ab}J+g_{ab}D,  \label{C.6}
\end{equation}%
\begin{equation}
\lbrack M_{ab},M_{cd}]=0,  \label{C.7}
\end{equation}%
\begin{equation}
p^{2}x_{a}=\varepsilon _{ab}Jp^{b}+\frac{1}{2}p_{a},  \label{C.8}
\end{equation}%
\begin{equation}
x\cdot p=-p\cdot x=\frac{1}{2},  \label{C.9}
\end{equation}%
\begin{equation}
p^{2}(x\cdot x)=J^{2}-\frac{1}{4},  \label{C.10}
\end{equation}%
\begin{equation}
\lbrack J^{2},p_{a}]=-p_{a}+2\varepsilon _{a}^{\ b}p_{b}J,  \label{C.11}
\end{equation}%
\begin{equation}
L_{ab}=M_{ab},\ \ S_{ab}=0,  \label{C.12}
\end{equation}%
\begin{equation}
p^{2}Q_{a}=-\frac{1}{2}J^{2}p_{a}+\frac{1}{2}p_{a}D^{2}-\varepsilon
_{ab}Jp^{b}D,  \label{C.13}
\end{equation}%
\begin{equation}
\lbrack p_{a},Q_{b}]=\varepsilon _{ab}J+g_{ab}D=[p_{a},q_{b}],  \label{C.14}
\end{equation}%
\begin{equation}
p^{4}[Q_{a},Q_{b}]=0,  \label{C.15}
\end{equation}%
\begin{equation}
R_{a}=q_{a}-Q_{a},  \label{C.16}
\end{equation}%
\begin{equation}
\lbrack R_{a},p_{b}]=0,\ \ [R_{a},R_{b}]=0,\ \ [R_{a},J]=0,  \label{C.17}
\end{equation}%
\begin{equation}
\lbrack D,R_{a}]=R_{a},\ \ [p^{2},R_{a}]=0,\   \label{C.18}
\end{equation}%
\begin{equation}
\lbrack R\cdot p,p_{a}]=0=[R\cdot p,J]=[R\cdot p,R_{a}],  \label{C.19}
\end{equation}%
\begin{equation}
\lbrack R\cdot R,p_{a}]=0=[R\cdot R,J]=[R\cdot R,R_{a}]=0,  \label{C.20}
\end{equation}
\end{center}

\end{document}